\documentclass[12pt]{iopart}
\usepackage{float}
\usepackage{graphicx}
\usepackage{epsfig}
\usepackage{bm}
\usepackage[flushleft]{threeparttable}
\usepackage{array,multirow}
\newcommand{\beq}{\begin{eqnarray}}
\newcommand{\eeq}{\end{eqnarray}}  
\begin{document}

\title[]{Electric-field-induced spin spiral state in bilayer zigzag graphene nanoribbons}

\author{Teguh Budi Prayitno}
 
\address{Department of Physics, Faculty of Mathematics and Natural Science, Universitas Negeri Jakarta, Kampus A Jl. Rawamangun Muka, Jakarta Timur 13220, Indonesia}
\ead{teguh-budi@unj.ac.id}
\vspace{10pt}

\begin{abstract}
We investigated the emergence of spin spiral ground state induced by the electric field in the bilayer zigzag graphene nanoribbons for the ferromagnetic edge states. To do that, we employed the generalized Bloch theorem to create flat spiral alignments for all the magnetic moments of carbon atoms at the edges within a constraint scheme approach. While the small ribbon width can preserve the ferromagnetic ground state, the large one shows the spiral ground state starting from a certain value of the electric field. We also pointed out that the spiral ground state is caused by the reduction of spin stiffness. In this case, the energy scale exhibits a subtle nature that can only be considered at the low temperature. For the last discussion, we also revealed that the spin spiral ground state appears more rapidly when the thickness increases. Therefore, we justify that the large ribbon width and large thickness can generate many spiral states induced by the electric field.            
\end{abstract}

\vspace{2pc}
\noindent{\it Keywords}: graphene nanoribbons, spin spiral, spin waves, spin stiffness
%
%
%
%

\section{Introduction}
The investigation on spin spiral ground state in the materials may be triggered by the experimental series by Tsunoda and Tsunoda $et$ $al.$ \cite{Tsunoda,Tsunoda1} when they observed a spiral ground state on $\gamma-$Fe, an fcc phase of iron. In the experiments, they stabilized the precipitates of $\gamma-$Fe at the low temperature in a fcc Cu matrix. So, the spiral ground state in the $\gamma-$Fe is a consequence of stabilizing the $\gamma-$Fe at the low temperature. After that, the theoretical discussions on this subject become popular and are still interesting. All the related authors found that the ground state of $\gamma-$Fe is very sensitive to the lattice parameter \cite{Uhl, Mryasov, Korling,Knopfle, Sjo}. 

The spiral (SP) state is a special case of helimagnetic state with a fixed cone angle \cite{Kurz}. Some papers reported that the magnetic domain wall can be portrayed by the SP formation to exploit physical features such as ferroelectricity or magnetoresistance for spintronic applications \cite{Sabirianov, Klaui, Stamps, Tokunaga,Wu}. Beside the bulk materials, further investigations also show that the SP state may emerge in the lower-dimensional materials, such as two-dimensional metal dihalides \cite{McGuire, Teguh3} and one-dimensional monoatomic chains \cite{Tows, Teguh}. So, we expect to find interesting physical properties induced by the SP formation in the low-dimensional materials. 

Intensive studies of exploiting the electronic and magnetic properties in the low-dimensional materials are initially driven by the discovery of graphene \cite{Novoselov1, Novoselov2, Geim}, a two-dimensional sheet of hexagonal carbon lattice. Previous studies show that graphene can be utilized for future devices, ranging from the electronic devices \cite{Liao, Schwierz}, optical devices \cite{Wang, Zhang}, to photonic devices \cite{Bonac, Bao}. Nevertheless, the magnetism in graphene can only be considered if its dimensionality is reduced into one-dimensional structure based on the theoretical study proposed by Fujita $et$ $al.$ \cite{Fujita} to become either the zigzag graphene nanoribbon (ZGNR) or the armchair graphene nanoribbon (AGNR). Here, we only consider the magnetic properties in the ZGNR.
        
It has been reported that the simplest way to achieve a metallic or an insulating property in the ZGNR is to arrange the magnetic moments of carbon atoms at the edges \cite{Sawada1}. Interestingly, this way can be used to create spin-wave excitations where the magnetic moments of carbon atoms along the edge are continuously rotated with the fixed cone angle. Previous studies exploit the spin-wave excitations in the monolayer ZGNR by using the supercell within the Hubbard model to investigate the critical temperature \cite{Yazyev,Yazyev1}, the lifetime of spin excitation \cite{Culchac1,Culchac2}, and the SP state \cite{Xie,Xiao}. For the latter, the other authors also observed the SP state in the monolayer ZGNR within the supercell approach by exploiting the non-equilibrium Green’s function method \cite{Zhang1,Liang} or by inserting the transition metal atoms \cite{Huang}. However, the wavevector, at which the SP state becomes a ground state, was not obviously elucidated.    

The above previous methods to find the SP state is not absolutely simple because the origin structure of ZGNR should be modified. Here, we present the simplest way but powerful to find the SP state in the ZGNR. Based on our previous results in the monolayer case \cite{Teguh2}, we continue to investigate the SP state in the bilayer ZGNR for the ferromagnetic edge states under the transverse electric field by using the generalized Bloch theorem (GBT). Due to the crystal structure, the explorations on the magnetic properties in the bilayer ZGNR are much more than those in the monolayer case \cite{Lima, Zhong, Sawada2, Teguh1}. 

There are two main reasons why the GBT is more powerful than the other approaches for investigating the SP state. The main advantage of using the GBT instead of the supercell is the lowest computational cost because it only requires the primitive unit cell. Even we do not need the additional atom to generate the SP state as in Refs. \cite{Liang, Huang}. Beside the efficiency,  the GBT can also give the SP ground state more accurately than the supercell when the SP state appears in the very small wavevector (long period) near $\Gamma$ point. Even when the supercell combined with the non-equilibrium Green’s function method or the Hubbard approach, it is still very difficult to determine the wavevector which gives the SP ground state.   

We found that the SP state appears due to the reduction of spin stiffness starting from a certain value of electric field, similar to the monolayer case \cite{Teguh2}. Here, the spin stiffness was calculated by using a least-square fit from the self-consistent total energy difference for a set of wavevectors. So, when the spin stiffness could be still calculated by this fit, the ground state is a ferromagnetic (FM) state. This means that there would be a phase transition from the FM state to the SP state as the spin stiffness can no longer be calculated by the fit. We also showed that this trend only occurs for the large ribbon width while the small one tends to preserve the FM state. Note that the SP state can only be induced if the initial state is the FM state whereas the antiferromagnetic (AFM) state is the most stable state in the multilayer ZGNR \cite{Sawada2}. This means that the SP state happens due to instability of the FM state under the electric field.
    
For the last session, we also discussed the influence of thickness on the SP state. When the thickness was taken into account, we observed a shift of electric field at which the SP state emerges for the first time. As the thickness increases, the electric field decreases, thus accelerating the emergence of SP state in terms of the electric field. For the larger ribbon width, the wavevector, at which the SP state emerges at a certain electric field, was larger than that for the smaller one for all the thicknesses. We also found that the SP state only emerges for the large ribbon width for all the thicknesses. This implies that not only the ribbon width but also the thickness can control the phase transition from the FM state to the SP state. Based on the results, we claim that the similar phase transition under electric field should also occur in the multilayer ZGNR.   
  
\section{Computational Method}
We used the OpenMX code \cite{Openmx} to perform the first-principles non-collinear calculations by implementing the GBT. Here, the wavefunction was expanded by the numerical linear combination of pseudo-atomic orbitals (LCPAO) as basis functions, which are produced within a confinement method \cite{Ozaki1, Ozaki2}. To do the efficient calculation, the norm-conserving pseudopotential \cite{Troullier} was used to represent the core Coulomb potential. For employing the GBT, the spiral wavevector $\mathbf{q}$ was inserted in the phase term of LCPAO written as \cite{Teguh4}
\beq
\psi_{\nu\mathbf{k}}\left(\mathbf{r}\right)&=&\frac{1}{\sqrt{N}}\left[\sum_{n}^{N}e^{i\left(\mathbf{k}-\frac{\mathbf{q}}{2}\right)\cdot\mathbf{R}_{n}}\sum_{i\alpha}C_{\nu\mathbf{k},i\alpha}^{\uparrow}\phi_{i\alpha}\left(\mathrm{\mathbf{r}-\tau_{i}-\mathbf{R}_{n}}\right)
\left(
\begin{array}{cc}
1\\
0\end{array} 
\right)\right.\nonumber\\
& &\left.+\sum_{n}^{N}e^{i\left(\mathbf{k}+\frac{\mathbf{q}}{2}\right)\cdot\mathbf{R}_{n}}\sum_{i\alpha}C_{\nu\mathbf{k},i\alpha}^{\downarrow}\phi_{i\alpha}\left(\mathrm{\mathbf{r}-\tau_{i}-\mathbf{R}_{n}}\right)\left(
\begin{array}{cc}
0\\
1\end{array}
\right)\right],\label{lcpao}
\eeq  
where the localized function $\phi_{i\alpha}$ is well defined within a cutoff radius as a boundary in the real space. Meanwhile, the flat spiral configuration was governed by the rotation of the magnetic moment $\mathbf{M}$ with a fixed cone angle $\theta$ 
\beq
	\mathbf{M}_{i}(t)=M_{i} \left(
\begin{array}{cc}
\cos\left(\varphi_{i}^{0}+\mathbf{q}\cdot \mathbf{R}_{i}+\omega_{\mathbf{q}} t\right)\sin\theta_{i}\\
\sin\left(\varphi_{i}^{0}+\mathbf{q}\cdot \mathbf{R}_{i}+\omega_{\mathbf{q}} t\right)\sin\theta_{i}\\
\cos\theta_{i}\end{array} 
\right). \label{moment}  
\eeq 

\begin{figure}[h!]
\vspace{-2mm}
\quad\quad\includegraphics[scale=0.6, width =!, height =!]{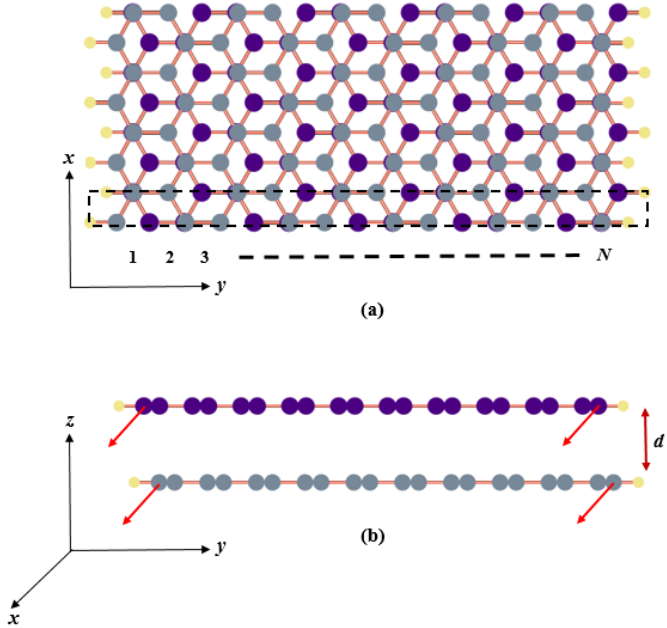}
\vspace{-2mm}
\caption{\label{model}(Color online) Crystal structure of AB-stacking bilayer ZGNR from top view (a) and side view (b). The initial FM state in (b) is set to create the flat spiral during the calculation. Here, the carbon and hydrogen atoms are depicted by the large and small spheres, respectively. Meanwhile, $d$ denotes the thickness, $N$ represents the ribbon width, and the dashed rectangle means the primitive cell.} 
\end{figure}

In this paper, we selected an AB-stacking primitive bilayer ZGNR due to its stability, as shown in Fig. \ref{model}. We set the experimental lattice parameter of 2.46 {\AA} in the \emph{x-} axis as a periodic lattice and thickness of 3.35 {\AA} from graphite in the \emph{z-} axis as a non-periodic direction. Then, we initially set an FM alignment of magnetic moments of carbon atoms at the four edges with $\theta=\pi/2$ to produce a flat spiral during the self-consistent calculation where the penalty functional was implemented to fix all the directions of magnetic moments \cite{Kurz}. In the calculation, the electric field $E$ was applied along \emph{y-} axis, parallel to the ribbon width $N$. Note that the applied $E$ can be used to induce the half-metallic property  within B3LYP exchange-correlation functional \cite{Rudberg, Kan}.   

For the detailed computation, two valence $s$-orbitals and two valence $p$-orbitals were specified for the carbon atoms while two valence $s$-orbitals plus one valence polarization $p$-orbital were set for the hydrogen atoms. At the same time, the cutoff radii for the carbon and hydrogen atoms are 4.0 Bohr and 6.0 Bohr, respectively. The non-collinear self-consistent calculation was then performed by using the Perdew, Burke, and Ernzerhof exchange-correlation functional \cite{Perdew} within $65 \times 1 \times 1$ $k$-point sampling and cutoff energy of 150 Ryd.  

\section{Results and Discussions}
\begin{figure}[H]
\vspace{-2mm}
\quad\quad\includegraphics[scale=0.5, width =!, height =!]{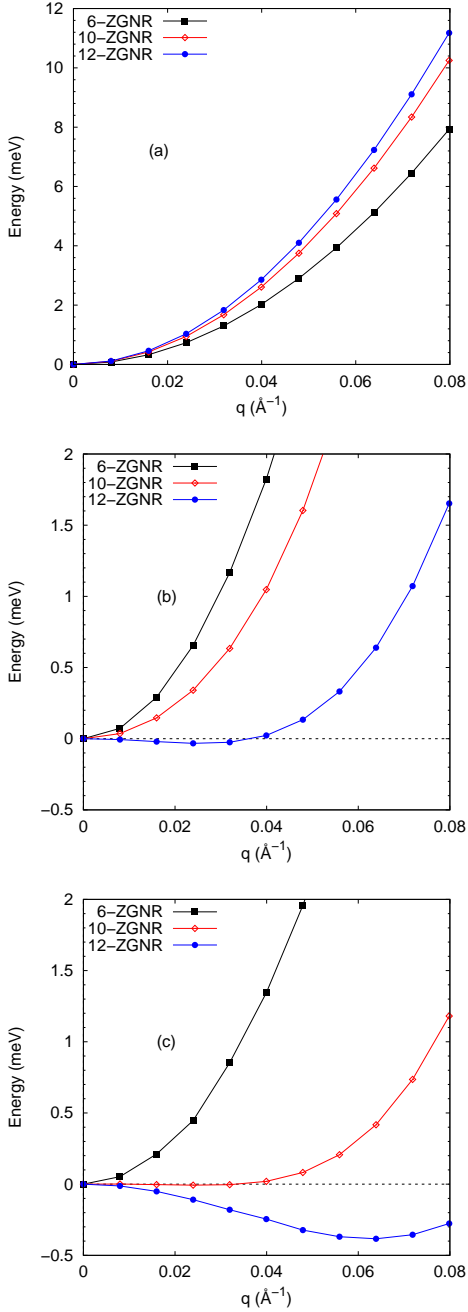}
\vspace{4mm}
\caption{\label{spiral} (Color online) Total energy difference $\Delta E=E(q)-E(q=0)$ as a function of $q$ along \emph{x-} axis (periodic direction) for $d=$3.35 {\AA} and $N=6,10,12$ with the applied electric field $E=0$ V/nm (a), $E=0.9$ V/nm (b), and $E=1.6$ V/nm (c) along \emph{y-} axis. Here, the SP ground states appear at $q=0.024$ for 12-ZGNR in (b), and at $q=0.024$ for 10-ZGNR and at $q=0.064$ for 12-ZGNR in (c).} 
\end{figure}

First, we consider the experimental thickness $d=$3.35 {\AA}. Figure \ref{spiral} shows the FM ground state ($q=0$) for the non-electric-field ($E=0$ V/nm) for all $N$ while the appearances of SP ground state occur at $E=0.9$ V/nm for 12-ZGNR, and at $E=1.6$ V/nm for 10-ZGNR. Meanwhile, no SP state is observed in 6-ZGNR for all $E$. Note that $E=0.9$ V/nm and $E=1.6$ V/nm are the initial values at which the SP state emerges for the first time for 12-ZGNR and 10-ZGNR, respectively. This means that the phase transition occurs from the FM state to the SP state after applying a critical $E$. 

\begin{figure}[H]
\vspace{-2mm}
\quad\quad\includegraphics[scale=0.6, width =!, height =!]{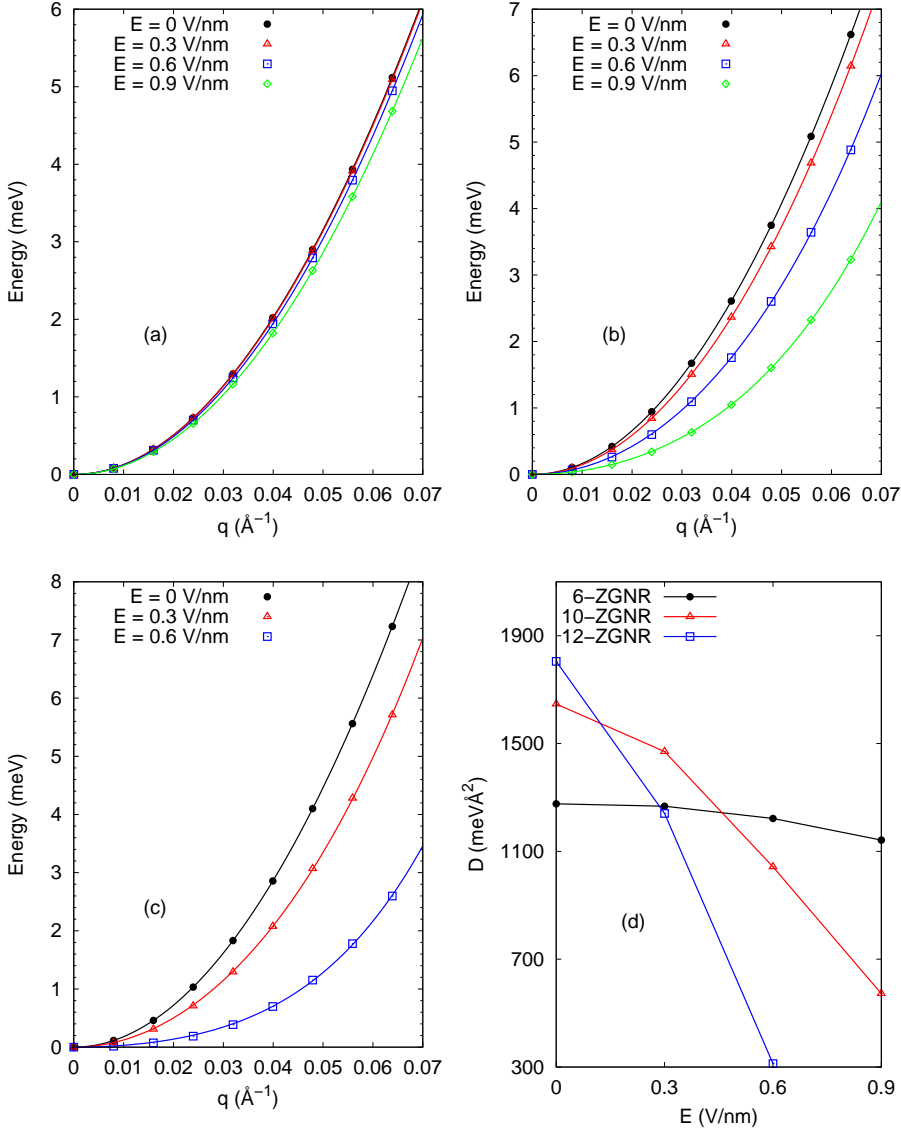}
\vspace{4mm}
\caption{\label{stiff} Total energy difference $\Delta E=E(q)-E(q=0)$ under $E$ for 6-ZGNR (a), 10-ZGNR (b), and 12-ZGNR (c) while the dependence of $D$ on $E$ is presented in (d) for $d=$3.35 {\AA}. Here, $D$ in (d) is obtained by fitting $\Delta E=D q^{2}(1-\beta q^{2})$ from (a-c).}
\end{figure}

When we check the spin stiffness $D$ with respect to $E$, we find that $D$ reduces as $E$ increases, as shown in Fig. \ref{stiff} for all $N$. This similar tendency was also reported by Rhim and Moon by applying the Hubbard Hamiltonian \cite{Rhim}. Note that $D$ is obtained by fitting the total energy difference through the equation $\Delta E=D q^{2}(1-\beta q^{2})$ in Figs. \ref{stiff}(a-c), where $q$ is defined in units of {\AA}$^{-1}$. In this case, $D$ can only be evaluated for the FM state. We also show that for $E=0$ V/nm, $D$ increases as $N$ increases, as shown in Fig. \ref{stiff}(d). However, at the same time in Fig. \ref{stiff}(d), the reduction of $D$ for the large $N$ is more rapid than that for the small $N$, similar to the monolayer case \cite{Teguh2}.   

The above trend can only be explained if the electron-electron interaction $J_{ij}$ among the edges only comes from the electron hopping. When $N$ increases, the energy increases, thus $D$ increases. At the same time, the electron hops easily from one edge to the other edges as $N$ decreases. This means that the large $N$ requires more energy to excite the spin waves than the small $N$ as the electron hops from one edge to the other edges. So, the value of $D$ is caused by the electron hopping, namely, the small $N$ gives the small $D$. Nevertheless, the smallest $N$, which has the strongest $J_{ij}$, tends to preserve the FM state when $E$ is applied. On the contrary, the large $N$ cannot preserve the FM state at a certain $E$, thus there is a limit value of $E$ to preserve the FM state. This implies that the strongest $J_{ij}$ in 6-ZGNR, even having the smallest $D$, can overcome the emergence of SP state due to the reduction of $D$. Therefore, we justify that the emergence of SP state is a consequence of reduction of $D$ and instability of FM state under $E$.        
\begin{figure}[h!]
\vspace{-2mm}
\quad\quad\includegraphics[scale=0.6, width =!, height =!]{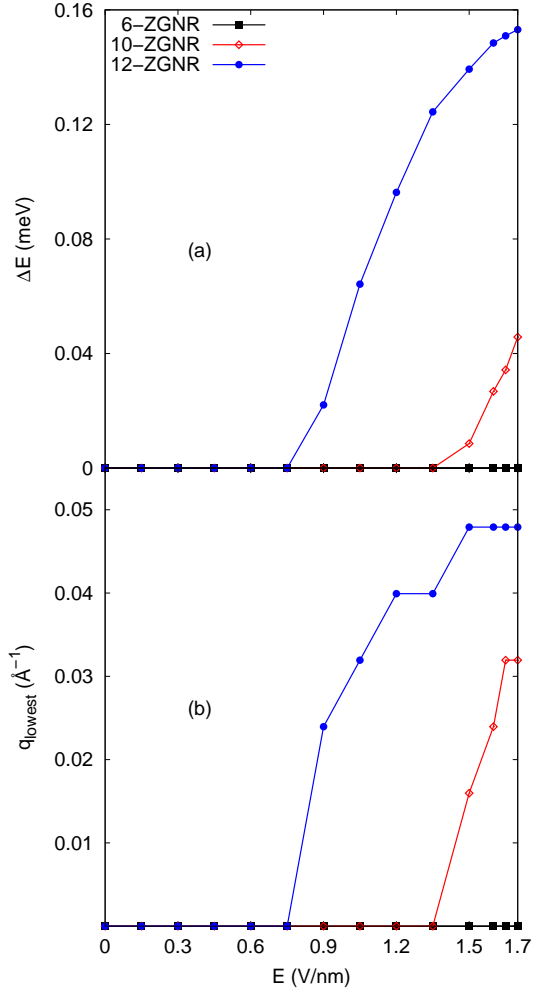}
\vspace{-9mm}
\caption{\label{2.95d} Dependence of total energy difference $\Delta E=E(q=0)-E(q=q_{\textrm{\scriptsize{lowest}}})$ on $E$ (a) and dependence of $q$ on $E$ (b) for $d=$2.95 {\AA}. Here, $q_{\textrm{\scriptsize{lowest}}}$ means $q$ having the lowest energy as the most stable state.}
\end{figure}	

\begin{figure}[h!]
\vspace{-2mm}
\quad\quad\includegraphics[scale=0.6, width =!, height =!]{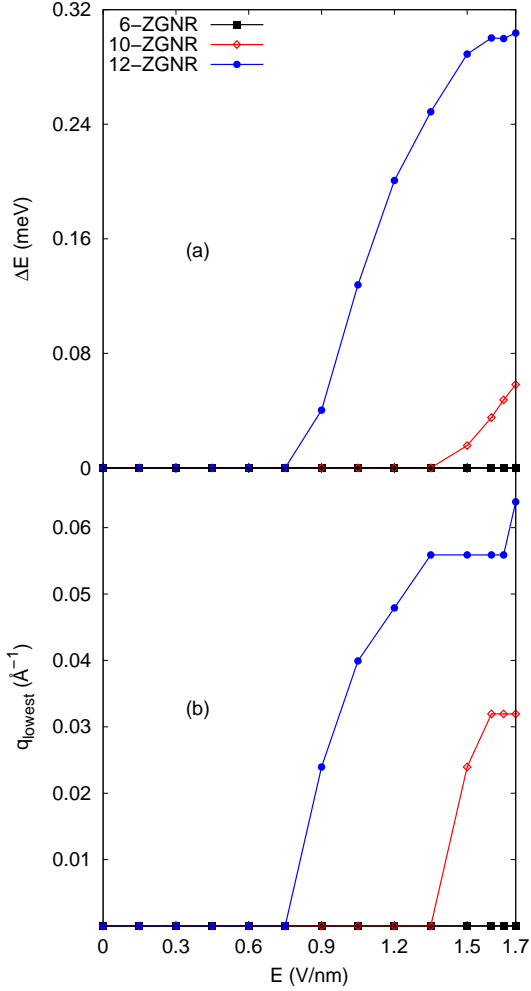}
\vspace{-9mm}
\caption{\label{3.15d} Dependence of total energy difference $\Delta E=E(q=0)-E(q=q_{\textrm{\scriptsize{lowest}}})$ on $E$ (a) and dependence of $q$ on $E$ (b) for $d=$3.15 {\AA}. Here, $q_{\textrm{\scriptsize{lowest}}}$ means $q$ having the lowest energy as the most stable state.}
\end{figure}
\begin{figure}[H]
\vspace{-2mm}
\quad\quad\includegraphics[scale=0.6, width =!, height =!]{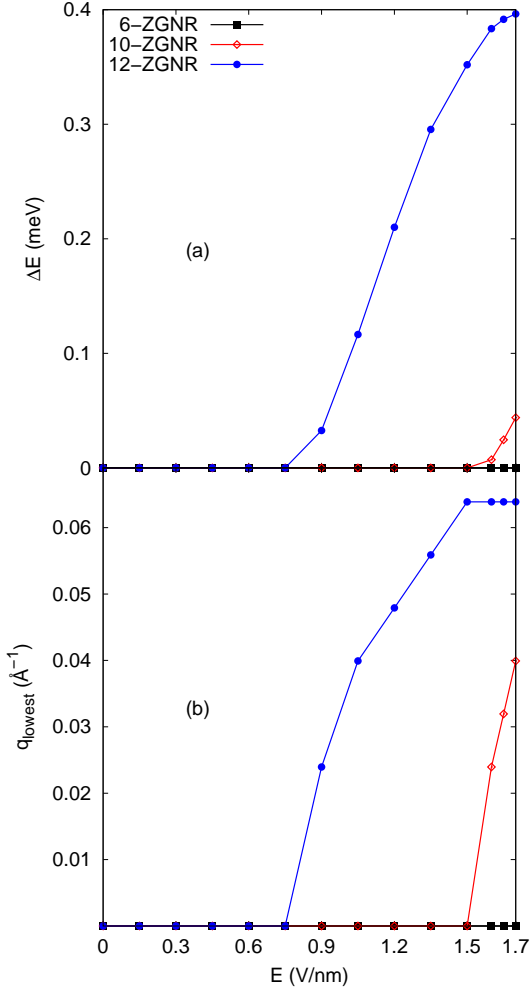}
\vspace{-9mm}
\caption{\label{3.35d} Dependence of total energy difference $\Delta E=E(q=0)-E(q=q_{\textrm{\scriptsize{lowest}}})$ on $E$ (a) and dependence of $q$ on $E$ (b) for $d=$3.35 {\AA}. Here, $q_{\textrm{\scriptsize{lowest}}}$ means $q$ having the lowest energy as the most stable state.}
\end{figure}

\begin{figure}[H]
\vspace{-2mm}
\quad\quad\includegraphics[scale=0.6, width =!, height =!]{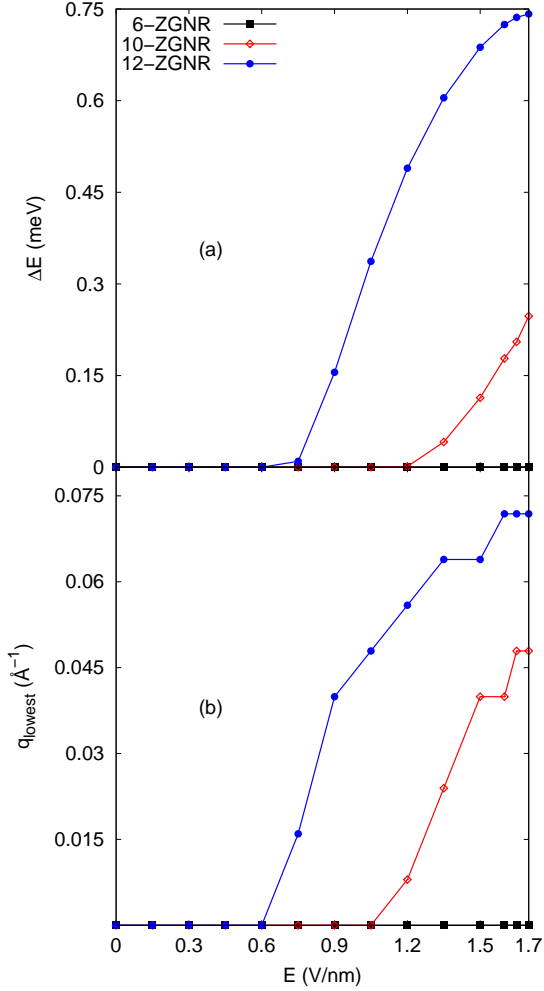}
\vspace{-9mm}
\caption{\label{3.55d} Dependence of total energy difference $\Delta E=E(q=0)-E(q=q_{\textrm{\scriptsize{lowest}}})$ on $E$ (a) and dependence of $q$ on $E$ (b) for $d=$3.55 {\AA}. Here, $q_{\textrm{\scriptsize{lowest}}}$ means $q$ having the lowest energy as the most stable state.}
\end{figure}

\begin{figure}[H]
\vspace{-6mm}
\quad\quad\includegraphics[scale=0.6, width =!, height =!]{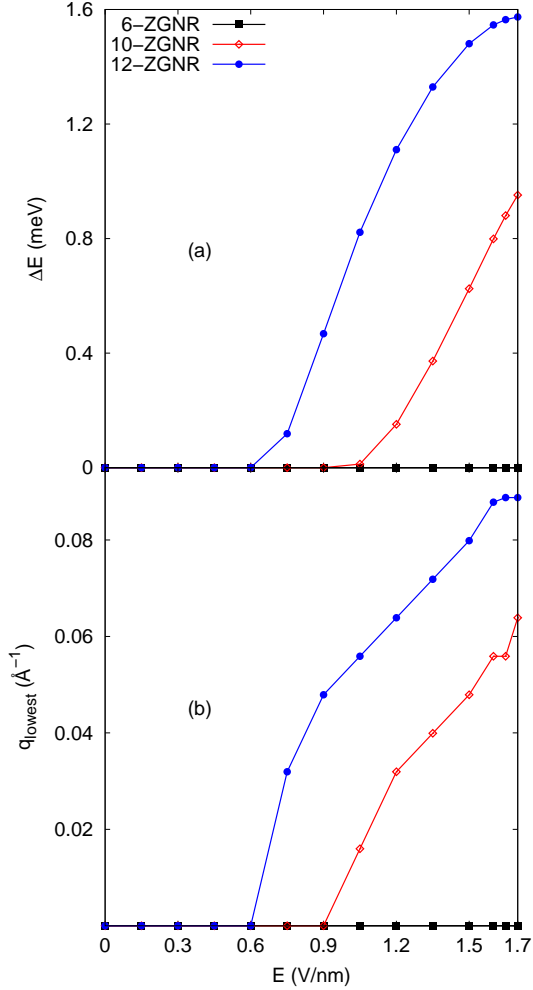}
\vspace{-9mm}
\caption{\label{3.85d} Dependence of total energy difference $\Delta E=E(q=0)-E(q=q_{\textrm{\scriptsize{lowest}}})$ on $E$ (a) and dependence of $q$ on $E$ (b) for $d=$3.85 {\AA}. Here, $q_{\textrm{\scriptsize{lowest}}}$ means $q$ having the lowest energy as the most stable state.}
\end{figure}

	
For the last discussion, we show the phase transition from the FM state to the SP state with respect to $E$ for several thicknesses $d$, as shown in Figs. \ref{2.95d}-\ref{3.85d}. For each $d$, we plot the total energy difference $\Delta E=E(q=0)-E(q=q_{\textrm{\scriptsize{lowest}}})$ as well as $q_{\textrm{\scriptsize{lowest}}}$ with respect to $E$, where $q_{\textrm{\scriptsize{lowest}}}$ is addressed to the most stable state. Our findings show that the 6-ZGNR still preserves the FM state as $E$ is applied. Contrarily, the 10-ZGNR and 12-ZGNR exhibit the SP state starting from a critical $E$, which also leads to a phase transition from the FM state to the SP state. We also observe that there are enhancements of $q_{\textrm{\scriptsize{lowest}}}$ and $\Delta E$ as $E$ increases, where the enhancements of the large $N$ are more rapid than those of the small $N$ for each $d$. However, each $d$ gives the different $q_{\textrm{\scriptsize{lowest}}}$ as well as the critical $E$.  

\begin{figure}[H]
\vspace{-2mm}
\quad\quad\includegraphics[scale=0.6, width =!, height =!]{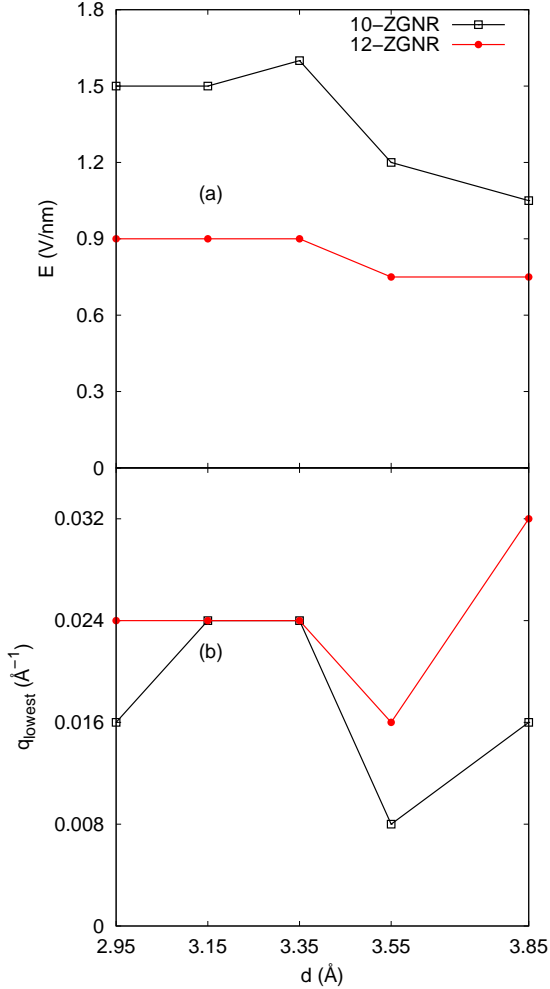}
\vspace{-9mm}
\caption{\label{tendency} Dependence of critical $E$ (a) and $q_{\textrm{\scriptsize{lowest}}}$ (b) on $d$ at which the SP state initially emerges.}
\end{figure}

For $d=2.95$ {\AA} as shown in Fig. \ref{2.95d}, the initial SP state is observed at $E=1.5$ V/nm and at $q_{\textrm{\scriptsize{lowest}}}=0.016$ for 10-ZGNR while the initial SP state in 12-ZGNR occurs at $E=0.9$ V/nm and at $q_{\textrm{\scriptsize{lowest}}}=0.024$. Next, when $d$ increases up to 3.15 {\AA} as shown in Fig. \ref{3.15d}, the initial SP state is still observed at $E=1.5$ V/nm but at $q_{\textrm{\scriptsize{lowest}}}=0.024$ for 10-ZGNR while the initial SP state in 12-ZGNR still occurs at $E=0.9$ V/nm and at the same $q_{\textrm{\scriptsize{lowest}}}=0.024$. For the experimental $d$ as shown in Fig. \ref{3.35d}, $E$ shifts to 1.6 V/nm for the initial SP state in 10-ZGNR at the same $q_{\textrm{\scriptsize{lowest}}}=0.024$ while the initial SP state in 12-ZGNR still occurs at $E=0.9$ V/nm and at the same $q_{\textrm{\scriptsize{lowest}}}=0.024$. When $d$ increases up to 3.55 {\AA} as shown in Fig. \ref{3.55d}, there are displacements of $E$ and $q_{\textrm{\scriptsize{lowest}}}$ for both 10-ZGNR and 12-ZGNR. The initial SP state in 10-ZGNR happens at $E=1.2$ V/nm and at $q_{\textrm{\scriptsize{lowest}}}=0.008$ while the initial SP state in 12-ZGNR occurs at $E=0.75$ V/nm and at $q_{\textrm{\scriptsize{lowest}}}=0.016$. The last one, for $d=3.85$ {\AA} as shown in Fig. \ref{3.85d}, $E$ shifts to 1.05 V/nm for the initial SP state in 10-ZGNR at $q_{\textrm{\scriptsize{lowest}}}=0.016$ while the initial SP state in 12-ZGNR still occurs at $E=0.75$ V/nm and at the different $q_{\textrm{\scriptsize{lowest}}}=0.032$. To give a better view, all those tendencies are then summarized in Fig. \ref{tendency} and table \ref{tabel}. In table \ref{tabel}, we add 8-ZGNR for the comparison with 6-ZGNR.

As immediately seen in Fig. \ref{tendency} (a), $E$ inclines to reduce when the SP state initially happens as $d$ increases. This trend is similar to the case of ribbon width $N$. Thus, $E$, which gives the SP state for the first time, only depends on the $J_{ij}$ between the two layers. By the same analogy in the case of $N$, the largest $d$ possess the weakest $J_{ij}$, thus the SP state occurs at the small $E$. Notice that this tendency also depends on $N$, where the given $E$ to achieve the SP state for the first time occurs earlier for the large $N$. On the contrary, we observe different trend for the $q_{\textrm{\scriptsize{lowest}}}$ when the SP state happens for the first time, as shown in Fig. \ref{tendency} (b). We find the $q_{\textrm{\scriptsize{lowest}}}$ inclines to decrease up to $d=3.55$ {\AA} and then increases. This means that $d=3.55$ {\AA} is a critical $d$ at which the SP state happens in the long wavelength (small $q$). Then, the short wavelength (large $q$) for the SP state appears in the large $d$. So, we justify that the ribbon width $N$ as well as the thickness $d$ give an important role to control the magnetic properties in the bilayer ZGNR due to the interactions between the magnetic carbon atoms at the edges.         

Now, we would like to give some comments on the obtained scale of energy. As shown in Figs. \ref{2.95d}-\ref{3.85d}, we can see that all the phase transitions from the FM state to the SP state happen in the order of few of meV. Compared to the thermal energy around 26 meV at the room temperature, this scale of energy should be sensitive to the thermal excitations. This implies that the resulting phase transitions under the electric field are subtle properties, which can only be investigated at the low temperature. This is due to a small magnetic moment ($\approx 0.3 \mu_{\textrm{\scriptsize{B}}}$) of each magnetic carbon atom at the edge in our calculations, which yields magnetic instability at the edge \cite{Kunstmann}. According to the previous reports \cite{Xiao, Liang, Huang}, the scale of energy of SP state can be increased up to the room temperature if the metal atoms are included. The existence of metal atom generates the charge transfer from the metal atom to the edge carbon atom, thus creating a strong bonding. This bonding will induce the robust magnetism, thus enhancing the scale of energy of SP state.       

Now, we would like to reveal some possible applications of SP state in the ZGNR. Previous papers reported that an SP configuration can form a domain wall that can be utilized for the spin transport in the spintronic devices by introducing a magnetic field \cite{Zhang1} or an atom doping \cite{Liang}. This spiral domain wall relies on the initial magnetization determined by the cone angle $\theta$. On the other hand, Zhang $et$ $al.$ \cite{Zhang3} stated that the SP configuration in the ZGNR induced by the Dzyaloshinskii-Moriya interaction may be utilized for the spin filters in the spintronic devices. This phenomenon can be realized when the ZGNR is grown on the topological insulator substrates. 
\begin{table}[htb]
\vspace{1 mm}
\caption{Initial $E$-induced spin spiral state with respect to $d$ for several $N$.}   
\centering 
\begin{tabular}{*{13}{c}} 
\hline 
\multirow{2}{5em}{\qquad$N$}& \multicolumn{5}{c}{$d$ ( {\AA})}\\
\cline{2-6}
&2.95& 3.15&3.35&3.55&3.85\\ 
\hline 
6& -& -&-&-&-\\ 
8& -& -&-&-&1.6 V/nm\\
10& 1.5 V/nm& 1.5 V/nm&1.6 V/nm&1.2 V/nm&1.05 V/nm\\
12& 0.9 V/nm& 0.9 V/nm&0.9 V/nm&0.75 V/nm&0.75 V/nm\\
\hline 
\end{tabular}
\vspace{0.01cm}
\label{tabel} 
\vspace{4 mm}
\end{table} 
\section{Conclusions} 
We prove the existence of the SP state induced by the electric field $E$ in the bilayer ZGNR by using the GBT. The observed wavevectors, at which the SP ground state appears, are absolutely small so that it is very difficult to generate the SP state using the supercell. The consequence of the small vector leads to a small energy scale that is very sensitive to the thermal excitations. In this case, the SP state induced by the electric field should be a subtle feature that cannot be observed at the room temperature. We also notice that the small ribbon width $N$ in the bilayer ZGNR preserves the FM ground state due to the strongest $J_{ij}$ while the large $N$ cannot maintain the FM ground state so that the ground state changes to the SP state, generating a phase transition from the FM state to the SP state.

We also show that not only the ribbon width $N$ but also the thickness $d$ can control the SP state as $E$ is applied. Here, we see the dependence of $E$ and $q_{\textrm{\scriptsize{lowest}}}$ on $d$ with different tendencies, except the small $N$. As $d$ increases, $E$ inclines to reduce while $q_{\textrm{\scriptsize{lowest}}}$ tends to decrease until a certain $d$ and start to increase. This condition also generates a phase transition from the FM state to the SP state. Note that all the energy scales for all $d$ are very small, too. Thus, the SP state for all $d$ are also subtle.  

We also believe that the SP states can also exist in the multilayer ZGNR as $E$ is introduced. They coexist with the FM edge states at the low temperatures. This is also due to the small magnetic moment of each edge carbon atom that yields the magnetic instability. In addition, as mentioned previously, since the most stable state in the ZGNR for any layer is always the AFM state, the FM state is unstable under $E$ for the large $N$ and large $d$. Therefore, the existence of SP state in the ZGNR under $E$ is the consequence of the instability of FM state.  
\section*{Acknowledgments}
All the detailed calculations were performed by using personal high computer at Universitas Negeri Jakarta. No funds are available in this research.

\end{document}